\documentclass[12pt,dvips]{article}
\usepackage{xtex}
\usepackage{graphicx}

\begin{document}

\begin{titlepage}

\setcounter{page}{1} \baselineskip=15.5pt \thispagestyle{empty}

\begin{flushright}
CPHT-RR037.0608 \\ 
\end{flushright}
\vfil

\begin{center}
{\LARGE Entanglement entropy of two dimensional systems and holography }
\end{center}
\bigskip\

\begin{center}
{\large Georgios Michalogiorgakis}
\end{center}

\begin{center}
\textit{Centre de Physique Th\'eorique, \'Ecole Polytechnique,
91128 Palaiseau, France
\\ Unit\'e mixte de Recherche 7644, CNRS
}
\end{center} \vfil


\noindent  In this note a new method for computing the entanglement entropy of a CFT holographically is explored.  It consists of finding a bulk background with a boundary metric that has the conical singularities needed to compute the entanglement entropy in the usual QFT definition.  An explicit calculation is presented for d=2.

\vfil

\end{titlepage}

\tableofcontents

\section{Introduction and summary}\label{INTRO}

One of the most interesting problems in theoretical physics surrounds black holes.  The nature of the entropy of the black holes, and its origin is still debated.  Amazingly, it only depends on the area of the horizon of the black hole
\eqn{BEKHAWK}{
S_{B.H.} = \frac{A}{4\pi G_{N}}\;,
} 
i.e. only on the geometric nature of the theory.  Another entropy, the geometric or entanglement entropy exists and exhibits similar geometric behavior.  Let us imagine a quantum field theory at some constant Euclidean time $t_{E}=t_{E,0}$ living on a manifold $\mathcal{M}$ and divide the manifold in two sub manifolds $\mathcal{A}$, $\mathcal{B}$.  The entanglement entropy measures how much the quantum states of the two regions are entangled.  Interestingly, a generic behavior for the entanglement entropy is 
\eqn{GENEE}{
S^{E.E} \sim \g \frac{\p \mathcal{A}}{\a ^{d-2}}+subleading \;terms,
}
where $\partial \mathcal{A}$ is the area of the boundary of $\mathcal{A}$, $\alpha$ is a UV cutoff and $d-1$ is the dimension of $\mathcal{M}$.  The constant $\g$ is generally non universal.  This is very similar to the black hole entropy and some identification of the UV cutoff with a Planck scale might reproduce \eno{BEKHAWK}.  Of course entanglement entropy has much broader interest.  It can be used in condensed matter and statistical physics as well.  Most of the calculations that have been done are for free theories or for theories with small interactions.  Of particular interest are CFT's because one expects conformality to fix some aspects of the result.

One would like to calculate the entanglement entropy of a given CFT, in the strong coupling limit.  Such a prescription exists \cite{Ryu:2006bv,Ryu:2006ef} for theories with gravitational duals.  It amounts to adding a term in the gravitational action of a $d-1$ dimensional surface term and finding the surface with the minimal area.  The entanglement entropy then is given by the Nambu-Goto like action of this $d-1$ dimensional surface.  Since this is computed in $AdS$ space we recover the correct scaling.  However, interestingly enough, in four dimensions there is a discrepancy between the anomalies one generically expects from QFT and the ones that are holographically realized \cite{Schwimmer:2008yh}, see \cite{Solodukhin:2008dh} for some counterarguments.

Quiet generally, on the QFT side the entanglement entropy is given by evaluating the  of the theory on a metric with a conical singularity, with a deficit angle $\d=2\pi \e$ \cite{Callan:1994py,Calabrese:2004eu}.  Then, the entropy is given by
\eqn{EEQFT}{
S^{E.E} =(\frac{\p}{\partial\e}-1)_{\e=0} W(E^\a)\;,
}
where $W(E^\a)$ is the effective action of the theory on the manifold with the conical singularity.  It should be noted that this formula is inspired by the analogous thermodynamics formula for computing the entropy given a partition function.
\eqn{STHERMO}{
S_{therm.}=-(\b\frac{\p}{\partial\b}-1)\log Z\;.
} 
 In the following we explore the extrapolation of this definition of entanglement entropy to the holographic duals of some CFTs.  In section \eno{EEQFTS}, a brief introduction to entanglement entropy and the methods used to compute it is given.  In section \eno{SINGBOUND} we extend the prescription for calculating the entanglement entropy to the holographic dual of the boundary CFT.  In section \eno{EXPLTWOD} an explicit calculation is carried for the $CFT_{2}/AdS_{3}$ system.  In section \eno{DISC} a brief summary of the results and some future directions are given. Appendix \eno{BTZ} contains a derivation for the BTZ case, illustrative of the general procedure used. 
 
\section{Calculation of Entanglement entropy in QFT}\label{EEQFTS}

In this section the earlier work of \cite{Callan:1994py,Calabrese:2004eu,Calabrese:2005zw,Holzhey:1994we} is briefly summarised.  Consider a QFT in a space that is artificially divided in two manifolds $A$, $B$.  Also, consider that the system is in a pure state $|\Psi\rangle$.  An observer that only has access to $A$ will not be able to measure the whole wavefunction but only the piece confined in his part of space.  So let us define a new density matrix
\eqn{DENSA}{
\r_{A}=\tr_{B} \r
} 
where $\r$  is the density matrix 
\eqn{DENSE}{
\r =|\Psi\rangle \langle\Psi|
}
and it is understood that the trace is taken over all the states of the Hilbert space of $\mathcal{B}$.  The entanglement entropy is just the von Neumann entropy of the reduced density matrix 
\eqn{EEVONN}{
S^{E.E.}_{\mathcal{A}} = - \tr \r_{\mathcal{A}} \log \r_{\mathcal{A}}\;.
}
For a product state we have $S^{E.E.}_{\mathcal{A}}=0$, while we expect the maximum value for a maximally entangled state.  The entanglement entropy has some interesting properties, such as $S^{E.E}_{\mathcal{A}}=S^{E.E}_{\mathcal{B}}$.  This explicitly shows that the entanglement entropy is non extensive.  Note that this equality is violated if the system is at finite temperature.  Another interesting property is strong subadditivity 
\eqn{EESTRONGSUB}{
S^{E.E.}_{\mathcal{A}}+S^{E.E.}_{\mathcal{A}'} \ge S^{E.E.}_{\mathcal{A}\cap \mathcal{A'}}+S^{E.E.}_{\mathcal{A}\cup \mathcal{A'}}\;.
}
The last property has been linked to an entropic analog of the Zamolodzikov's c theorem \cite{Zamolodchikov:1987ti} in two dimensions \cite{Casini:2004bw,Casini:2003ix}.

Now, let us move on to properly calculating the entanglement entropy.  For simplicity let us consider a bosonic theory with a complete set of commuting observables $\{\hat{\phi}(x) \}$, whose eigenvalues are given by $\{ \phi(x)\}$.  The evolution of the theory is governed by a Hamiltonian $\hat{\mathcal{H}}$ and the density matrix at some inverse temperature $\b$ is given by 
\eqn{DENSMATRIXTHE}{
\r = (\{\phi(x'')''\},\{\phi(x')'\})=\frac{\langle \{\phi(x'')''\}|e^{-\b \hat{\mathcal{H}}}|\{\phi(x')'\}\rangle}{Z(\b)}\;,
} 
where $Z(\b)$ is the partition function.  This also has an expression as a path integral
\eqn{DENSPATH}{
\r = \frac{1}{Z} \int \left[ d\phi(x,t) \right] \prod_{x} \d(\phi(x,0)-\phi(x')')\prod_{x}\d(\phi(x,\b)-\phi(x'')'')e^{-S_{E}}\;,
} 
where we have introduced the Euclidean action $S_{E}$.  The normalization is such that $\tr \r =1$.  Now consider a single interval $\mathcal{A}=(u,v)$.  A similar expression can be written for the reduced density matrix $\r_{\mathcal{A}}$, but with sewing together only the points which do not belong to $\mathcal{A}$.  Of course, this leaves open cuts in $(v,u)$ for the Euclidean time $t_{E}=0$.  The desired trace $\tr \r^{n}_{\mathcal{A}}$ is computed by making $n$ copies and sewing them together along the cuts.  If we denote by $k$ the $k$-th copy then the sewing conditions are 
\eqn{SEWPHI}{
\phi_{k}'(x)=\phi_{k+1}''(x),\quad \phi_{n}'(x)=\phi''_{1}(x),\quad x\in \mathcal{A}\;.
} 
This way, we define a path integral on an $n$-sheeted manifold $\mathcal{R}_{n}$, $Z_{n}(\mathcal{A})$.  It is understood that the primed fields are evaluated at a time $t_{E}=0^{-}$, while the double primed are evaluated at an infinitesimal positive time $t_{E}=0^{+}$. It is more convenient to introduce a new coordinate $w$
\eqn{DEFW}{
w= x+i t_{E}\;,
}
and then the sewing conditions become
\eqn{GOTSEW}{
\phi_{k}(e^{2\pi i}(w-u)) = \phi_{k+1}(w-u),\quad \phi_{k}(e^{2\pi i}(w-v))=\phi_{k-1}(w-v)\;.
}
The trace of the density matrix becomes
\eqn{GOTRACEN}{
\tr \r_{\mathcal{A}}^{n} = \frac{Z_{n}(\mathcal{A})}{Z_{1}^{n}}\;.
}
In \cite{Calabrese:2004eu}, whose discussion we have closely followed, some arguments are given that the above quantity is analytic for $Re (n) >1$ and that it's derivative with respect to $n$ exists and is analytic in the same region.  Moreover the derivative at $n=1$ precisely gives the entanglement entropy 
\eqn{EERN}{
S^{E.E.}_{\mathcal{A}} = - \lim_{n\rightarrow 1}\frac{\p}{\p n} \frac{Z_{n}(\mathcal{A})}{Z_{1}^{n}}\;.
}

In order to practically compute the entanglement entropy, Calabrese and Cardy in \cite{Calabrese:2004eu} propose as a first step a conformal transformation that transforms the initial space $\mathcal{R}_{n}$ to $\mathcal{C}$. Then, it is argued that the ratio of the two partition functions of \eno{EERN} is the same as correlation functions arising from the insertion of primary scaling operators at the points $v,u$.  We will take a different route and argue that the entanglement entropy is given by computing the effective action of the CFT $W_{\mathcal{C}}(n)$ on $\mathcal{C}$ and taking the derivative with respect to $\e=n-1$, as in \eno{EEQFT}.  Note that $\mathcal{C}$ inherits the metric from the conformal transformation from $\mathcal{R}_{n}$.  

\begin{figure}[htp]
\centering
\includegraphics[width=8cm]{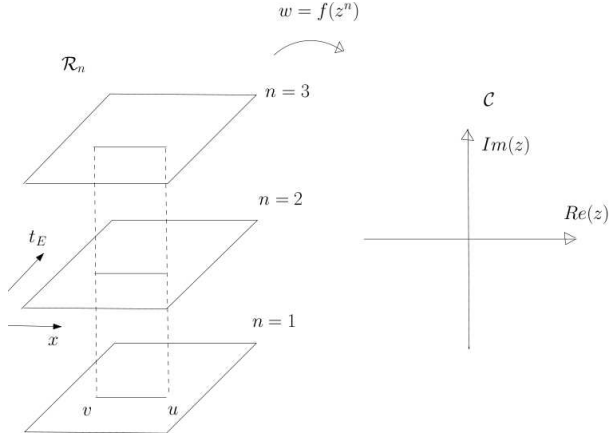}
\caption{A conformal transformation that takes us from $\mathcal{R}_{n}$ to $\mathcal{C}$ for $n=3$.  For the calculation of the entanglement entropy of an interval $(v,u)$ the required transformation is given by $ \left(\frac{w-u}{w-v} \right)^{1/n} $.  It is understood that when circling around one endpoint, one ascends to the next sheet or descends to the previous one.  More specifically \eno{GOTSEW} suggests that circling around $x=u$ one ascends to the next sheet, while circling around $x=v$ one descents to the previous one.}\label{MyFigure}
\end{figure}

Since we now have a CFT on a curved background, it is expected that the entanglement entropy will be dominated by the conformal anomaly.  The variation of the entropy with respect to the length of $\mathcal{A}$ is similar to a Weyl variation and so, for $L=|u-v|$ 
\eqn{WEYLEE}{
L\frac{d}{d L} S^{E.E}_{\mathcal{A}} \sim \int \langle T^{\m}_{\m} \rangle \sim \int R \;.
} 

Let us examine this process for a simple case.  Specifically, if one starts from a single interval on a infinite space and zero temperature, the process of "uniformising" is achieved by 
\eqn{ZUNIF}{
z= \left(\frac{w-u}{w-v} \right)^{1/n}\;.
}
If the initial manifold is at a temperature $T=\frac{1}{\b}$ then we first need to transform
\eqn{}{
w'=\frac{\b}{2\pi}\log w
} 
and then uniformise as in the previous example.  A similar procedure is applied if the manifold has space periodicity $R$ and is at zero temperature, replacing $R\rightarrow i \b$ in the previous formulas.  A schematic of this transformation is presented in figure \eno{MyFigure}.

The result of the whole procedure can be summarised by the value of the entanglement entropy for some specific cases.  For the single interval we have 
\eqn{EESINGLE}{
S^{E.E.}_{\mathcal{A}} = \frac{c}{3}\log\frac{L}{\a} +c_{1}\;, }
where $\a$ is the UV cutoff and the finite part $c_{1}$ is depended on the details of the theory, it is not universal.  Similarly for finite temperature we have 
\eqn{EEFINITET}{
S^{E.E.}_{\mathcal{A}} = \frac{c}{3} \log \left(\frac{\b}{\pi \a}\sinh(\frac{\pi L}{\b})\right) +c_{2} \;,
}  
where $c_{2}$ is also a non universal quantity.
\section{Extension to holographic theories}\label{SINGBOUND}

The AdS/CFT correspondence \cite{Maldacena:1997re} teaches us that there is a duality between string theory living in asymptotically AdS spaces and a conformal field theory living on the boundary.  The question of the exact correspondence between quantities in the CFT and quantities of the string theory is an interesting one.  Fortunately it has been known from the early days of the correspondence that the partition function of the CFT is related to the string theory action \cite{Gubser:1998bc,Witten:1998qj} (for a review of the correspondence see \cite{Aharony:1999ti}).  Explicitly 
\eqn{BULKBOUNDCORR}{
Z_{CFT}(h) = Z_{S}(h)  
}       
where $Z_{CFT}(h)$ is the partition function of the CFT on a manifold with conformal structure $h$.  $Z_{S}$ is the exponential of the action of string theory integrated over all metrics that have a double pole on the boundary and induce the given conformal structure.  In the limit of large N and small curvatures the latter becomes the effective low energy supergravity action.
\eqn{SUGRAS}{
Z_{S}(h)=e^{-S_{sugra}(g)}\;,
} 
with $g$ a solution of Einstein's equations with the expected boundary behavior.

It is straightforward to extend the definition of the entanglement entropy to the bulk theory.  As has been  observed before \cite{Schwimmer:2008yh,Fursaev:2006ih}, we argue that the following procedure should be followed in order to calculate the entanglement entropy holographically:

\begin{itemize}
 \item  Define the theory on the boundary with a metric that has the required conical singularity $g_{\e,\partial \mathcal{M}}$.
 \item  Find a bulk metric with the desired asymptotic behavior.  One can proceed and holographically reconstruct the bulk metric, as is explained for example in \cite{Skenderis:1999nb,deHaro:2000xn}.  Choosing Fefferman-Graham coordinates \cite{GrahamFeff}, each term of the series in $\r$ can be determined algebraically from the previous one.   This is a well defined procedure for an integer $n$ and the result is analytically continued to arbitrary $n$.  It could be argued that this step is not well defined for arbitrarily small $\e=n-1$, since in higher orders in $\r$ one would introduce more severe singularities.  
 \item Finally one needs to evaluate the bulk action
\eqn{BULKACTION}{
S_{bulk} = \frac{1}{16 \pi G_{N} }\int_{\mathcal{N}}\sqrt{-det(g)}\left(R+2\L\right) +\frac{1}{8\pi G_{N}} \int_{\partial \mathcal{N}}\sqrt{-det(h)}\Th\;. 
}
 to first order in the deficit angle $\e$ and compute the entanglement entropy from 
\eqn{EEBULK}{
S^{E.E.} = \left( \frac{\partial}{\p \e} -1\right)_{\e \rightarrow 0} S_{bulk}\;.
}
$R$ is the Ricci scalar and $\Th$ is the extrinsic curvature on the boundary.
\end{itemize}

Additionally to this straightforward attack of the problem, an alternative has been proposed.  It has been argued recently \cite{Ryu:2006bv,Ryu:2006ef} that the entanglement entropy can be computed holographically by minimizing a $d-1$ dimensional surface that has as a boundary  $\partial A$.  Some arguments are given in \cite{Fursaev:2006ih} that the prescription of the previous section reproduces the Ryu-Takayanagi proposal.  However,  there is an "ad hoc" addition to the bulk action.  This proposal gives the expected results for $d=2$ theories and proposes reasonable entanglement entropies for higher dimensional CFT's or generically theories with gravitational duals.  It has been used for a variety of very interesting computations, from suggesting that entanglement entropy can be used for an order parameter for confinement/deconfinement \cite{Nishioka:2006gr,Klebanov:2007ws} to examining aspects of black holes \cite{Azeyanagi:2007bj}, see also \cite{Hubeny:2007re,Hubeny:2007xt,Pakman:2008ui,Barbon:2008ut,Azeyanagi:2007qj,Headrick:2007km,Cadoni:2007vf,Cadoni:2007vf,Hirata:2006jx} for similar lines of thought.   

The analytic structure of both conformal and Graham-Witten anomalies \cite{GrahamFeff} for the Ryu-Takayanagi proposal in $d=4$ dimensions has been examined in \cite{Schwimmer:2008yh}, note also the counterarguments of \cite{Solodukhin:2008dh}.\footnote{I am grateful to A. Schwimmer for discussions regarding the issues of anomalies.}  There, a mismatch has been found between the holographic Ryu-Takayanagi prescription and the usual QFT entanglement entropy calculation. A possible resolution could be that the replica trick fails in higher dimensions.  Another one could be that the Ryu-Takayanagi prescription calculates some other Wilson loop type of observable.  The proposal of this paper evades these problems, since the anomalies will match by construction.  In a sense, the anomalies of the boundary theory dictate the anomalies of the bulk.  The most interesting cases would be to examine the results of $d=4$ and see whether there is agreement between the two approaches.  We will examine here only the case of $d=2$ as an exercise of implementing the approach.  Unfortunately the technical difficulty of finding bulk solutions increases with the dimension.

Another nice application of the formulation of this section is that it is inherently easy to tackle time dependent backgrounds.  As noted in \cite{Calabrese:2005in}, in two dimensions it is possible to derive the entanglement entropy of a time dependent manifold.  In \cite{Hubeny:2007xt}, the prescription of \cite{Ryu:2006bv,Ryu:2006ef} has been extended to these situations.  In principle there should be a direct connection between that approach and the proposal followed here.  Finally, one should also note that it is possible to follow these three steps for an arbitrary integer $n$.  Then one should use in the last step the derivative with respect to $n$ to derive the Tsallis entropy \cite{Tsallis:1987eu}.  
\eqn{TSALISE}{
S_{Ts.} = \frac{\tr \r_{\mathcal{A}} ^n -1}{n-1} \;.
}

\section{Calculation in two dimensional CFTs and their holographic duals}\label{EXPLTWOD}

In this section we explore the definition of \eno{EEBULK} specifically for two dimensional CFTs.  Generically one expects such CFTs to be dual to a theory in an $AdS_{3}\times \mathcal{K}$ background.  We will ignore the complexities that arise from including higher dimensions and will only deal with asymptotically $AdS_{3}$ geometries.

One way to circumvent the discrepancy in the anomalies between the holographic prescription and the QFT calculation is to try to find a bulk solution that asymptotically aproaches the metric with the required conical singularity in the boundary.  In this way one has a matching of anomalies on the two sides ``by construction``.  In order to explore this possibility let us work in the case where the boundary theory is a CFT in 2 dimensions and the bulk solution is an asymptotically $AdS_3$ spacetime.  In this case there is no problem in the anomaly matching since it is almost trivial, but we would like to see whether this direct approach makes sense.  We are working in the Fefferman-Graham coordinates and the metric can be expanded as
\eqn{METRICEXPAND}{
ds^2 = \frac{\ell^2 d\r^2}{4\r^2}+\frac{g_{ij}(x^k,\r)dx^i dx^j}{\r}\;,
}  
where the $d$ dimensional metric, itself can be expanded as 
\eqn{TWOMETRICEXP}{
g_{ij} = g_{0,ij}+\r g_{2,ij}+ \r^2 g_{4,ij}+\mathcal{O} \left( \r^{3} \right) \;.
}
The case of $d=2$ is of particular interest.  As explained in \cite{Skenderis:1999nb} the series terminates at $g_4$.  The whole metric is given by
\eqn{GOTMETRIC}{
g=\left(1+\frac{\r}{2}g_{2}g_{0}^{-1} \right)g_{0}\left(1+\frac{\r}{2}g_{0}^{-1}g_{2}\right)\;.
}
One only needs to determine $g_2$.  Let us quickly review how this is done \cite{Skenderis:1999nb}.  The second component of the metric $g_2$ is given by
\eqn{GOTGTWO}{
g_{2,ij}= \frac{\ell^2}{2}\left(R_0 g_{0,ij}+T_{ij} \right)\;,
}  
where $T_{ij}$ is a symmetric traceless tensor that is given by 
\eqn{GOTT}{
T_{ij}=\frac{1}{2}\nabla_{i}\phi\nabla_{j}\phi +\nabla_{i}\nabla_{j}\phi-\frac{1}{2}g_{0,ij}\left(\frac{1}{2}(\nabla\phi)^2+2\Box \phi\right)\;.
}
The scalar field $\phi$ satisfies 
\eqn{PHIEOM}{
\Box \phi=R_0\;.
} 
It should be noted that it is allowed to add to $T_{ij}$ the stress energy tensor of arbitrary conformal matter, or put differently a traceless, covariantly conserved tensor.  This fact should be taken into account when constructing bulk solutions.  Following the prescription, we are interested in finding a solution that has a boundary metric with a conical singularity
\eqn{BOUNDMETRIC}{
ds^2=g_{0,ij}dx^{i}dx^{j}=(z\bar{z})^{n-1}f'(z^{n})f'(\bar{z}^{n})dz d\bar{z}
}
and in particular in taking the limit $n-1=\e\rightarrow 0$.  Since it is natural to uniformise the interval $A$ we allow for a conformal transformation that takes care of this.  In the following, we will only examine the case of small $\e$ and expand all quantities in an $\e$ series.  It is very convenient to write 
\eqn{METRICEPSI}{
ds^{2}= g_{0,ij}dx^{i}dx^{j}=f'(z)f'(\bar{z})\left(1-\e G(z,\bar{z})\right)dz d\bar{z}
}
where
\eqn{GOTG}{
G(z,\bar{z})=-\log(z\bar{z}) -\frac{z\log z f''(z)}{f'(z)}-\frac{\bar{z}\log \bar{z} f''(\bar{z})}{f'(\bar{z})}\;.
}
We expect the Riemann and Ricci tensor to have derivatives of $G(z,\bar{z})$.  It is straightforward to calculate that 
\eqn{DERG}{
\frac{\partial^2 G}{\partial z\partial \bar{z}} = -\frac{\partial^2 }{\partial z\partial \bar{z}} \log z\bar{z} = -2\pi \d^{2}(z,\bar{z})\;.
}
In Einstein's equations of the bulk solution one also finds other derivatives of $G$, such as $G^{2,1}$, $G^{1,2}$ and so on.  These should be interpreted as derivatives of Dirac's delta function. 
\eqn{DERDIR}{
\int dz d\bar{z} G^{2,1}(z,\bar{z})F(z,\bar{z}) = - \int dz d\bar{z} G^{1,1}(z,\bar{z})F^{1,0}(z,\bar{z}) =  2\pi F^{1,0}(0,0)\;.
}
For the metric \eno{METRICEPSI} the Ricci scalar is easily computed to be 
\eqn{GOTZERORSCALAR}{
R_0= \e \frac{\partial^2_{z, \bar{z}} G(z,\bar{z})}{f'(z)f'(\bar{z})}\;.
} 
and \eno{PHIEOM} gives
\eqn{GOTPHIEQN}{
\frac{\partial^2_{z, \bar{z}} \phi(z,\bar{z})}{f'(z)f'(\bar{z})}=  \e \frac{\partial^2_{z, \bar{z}} G(z,\bar{z})}{f'(z)f'(\bar{z})} ;.
} 
with solution 
\eqn{GOTPHI}{
\phi(z,\bar{z}) = \phi_{backgr.}(z,\bar{z})+ G(z,\bar{z})\;.
}
The background metric satisfies 
\eqn{HOIBACK}{
\Box \phi_{backgr.} =0
}
and allows for a non trivial asymptotically AdS metric.  In appendix \eno{BTZ} it is demonstrated how, for example the BTZ black hole solution is generated.  The next step is to calculate $g_{2}$ to order $\e$.  For simplicity let us write
\eqn{GOTPHIBACK}{
\phi_{backgr.}(z,\bar{z})=Az +\bar{A}\bar{z} +\Gamma\;.
} 
The second term in the metric expansion is then found to be 
\eqn{GOTGTWOE}{
g_{2,ij} = \left(\begin{array}{cc} \frac{1}{4}A^2 \ell^2 +\e \frac{\ell^2}{2}\left(A\partial_{z}G +\partial^{2}_{z,z} G \right) & \e \frac{\ell^2}{2}\partial^{2}_{z,\bar{z}}G \\ \e \frac{\ell^2}{2}\partial^{2}_{z,\bar{z}}G & \frac{1}{4}\bar{A}^2 \ell^2 +\e \frac{\ell^2}{2}\left(\bar{A}\partial_{\bar{z}}G +\partial^{2}_{\bar{z},\bar{z}} G \right) \end{array}  \right)\;.
}
Finally one is ready to write down the whole bulk solution metric 
\eqn{GOTWHOLEMETRICA}{
ds^2 =\frac{\ell^2 d\r^2}{4 \r^2} +\frac{ g_{ij}dx^{i}dx^{j}}{\r}=\frac{\ell^2 d\r^2}{4 \r^2}+\frac{\left(g_{0,ij} +\r g_{2,ij}+\r^2g_{4,ij}\right)dx^{i}dx^{j}}{\r}
}
where $g_0$ is given in \eno{METRICEPSI}, $g_2$ in \eno{GOTWHOLEMETRICA} and $g_4$ has the more complicated form
\eqn{GOTFOURMETR}{
g_{4,ij} = &\frac{\ell^4}{128f'(z)f'(w)}\cdot \cr
& \cdot \left(  
 \begin{array}{cc} 
 4 A^{2}\e \partial^{2}_{z\bar{z}} G & A^2\bar{A}^2(1+\e G)+2 \e A\bar{A}^2 \partial_{z} G + 2\e A^2 \bar{A} \partial_{\bar{z}} G +2\e A^2 \partial^{2}_{\bar{z}}G+2 \e \bar{A}^2 \partial^{2}_{z} G\\
  g_{4,z\bar{z}}       & 4\bar{A}^{2}\e\partial^{2}_{z\bar{z}}G
 \end{array}
\right)\;,
}
where the off-diagonal element has not been written twice.  It is a straightforward but painstaking exercise to verify that \eno{GOTWHOLEMETRICA}-\eno{GOTFOURMETR} are a solution to Einstein's equations to order $\e$
\eqn{EINSBUKSOL}{
R_{\m \n} -\frac{1}{2}Rg_{\m\n} = \L g_{\m\n} +\mathcal{O}(\e^2)\;.
}

\subsection{Calculating the holographic entanglement entropy}\label{SECBULKACTION}

Now all the tools necessary for evaluating the entanglement entropy are in order.  The required action  \eno{BULKACTION}
\eqn{BULKACTIONA}{
S_{bulk} = \frac{1}{16 \pi G_{N} }\int_{\mathcal{N}}\sqrt{-det(g)}\left(R+2\L\right) +\frac{1}{8\pi G_{N}} \int_{\partial \mathcal{N}}\sqrt{-det(h)}\Th\;.
}
We only need to evaluate this action to first order in $\e$.  Since Einstein's equations are satisfied to order $\e$ we have 
\eqn{EINSAT}{
R_{\m \n} -\frac{1}{2} R g_{\m \n} = \L g_{\m \n} +\mathcal{O}(\e^{2})\Rightarrow R=-\frac{6}{\ell^2} +\mathcal{O}(\e^2) \;.
}
Then we need to calculate the determinant and the extrinsic curvature term.  The results presented here are in a Laurent series in $\e$ and $\r$.  Since 
\eqn{GOTSQRDET}{
\sqrt{-det(g)}=\frac{\ell f'(z)f'(\bar{z})(1-\e G(z,\bar{z}))}{\r^2} + \e \frac{\ell^3 G^{1,1}(z,\bar{z})}{4\r }+ \mathcal{O}(\e^2,\r^0)\;,
} 
the first term evaluates to 
\eqn{GOTFIRSTERM}{
S_{1} =  -\frac{1}{2\pi G_{N}} \int_{\p \mathcal{N}} d z d\bar{z}\left(\frac{f'(z)f'(\bar{z})(1-\e G(z,\bar{z}))}{\ell \r_{min}} +\e \frac{\ell}{4}G^{1,1}(z,\bar{z}) \log\frac{\r_{max}}{\r_{min}}\right) + \mathcal{O}(\e^2,\r_{min}^{1})\;.
}
The finite terms are known but are very complicated to write down explicitly for the general case here.  In order to calculate the extrinsic curvature part of the action one needs to find the normal to the surface $\r=const.$, $n^{\m}$ and then the extrinsic curvature is given by
\eqn{GOTEXTRINSIC}{
\Th^{\m \n}=-\frac{1}{2}\left(\nabla^{\n}n^{\m}+\nabla^{\m}n^{\n} \right)\;.
}
For the specific case of the boundary being the surface $\r=const.$ the unit normal is 
\eqn{GOTNORMAL}{
n^{\m}=\frac{2\r}{\ell}\d^{\m}_{\r}\;.
}
It is easy to calculate the extrinsic curvature and it turns out to be
\eqn{GOTEXTSOL}{
\Th_{ij}=-\frac{\r}{\ell}\p_{\r}\frac{g_{0,ij}+\r g_{2,ij}+\r^2 g_{4,ij}}{\r}=\frac{g_{0,ij}-2\r^2g_{4,ij}}{\ell \r}\;.
}
We take the boundary of $\mathcal{N}$ to be at a finite $\r=\r_{min}$, and the second part of the action, to first order in $\e$ becomes
\eqn{GOTEXTACTION}{
S_{2}=\frac{1}{2\pi G_{N}}\int_{\p \mathcal{N}}dzd\bar{z}\left(\frac{1}{\ell \r_{min}} f'(z)f'(\bar{z})(1-\e G(z,\bar{z}) +4\ell \e G^{1,1}(z,\bar{z})\right) +\mathcal{O}(\e^2,\r_{min}^{1}) \;.
}
Combining the two terms and using \eno{BULKACTIONA} 
\eqn{GOTSBULK}{
S_{bulk} = -\frac{1}{8\pi G_{N}} \int_{\p \mathcal{N}} dz d\bar{z} \e \ell G^{1,1}(z,\bar{z}) \left(\log\frac{\r_{max}}{\r_{min}}+\frac{1}{4}\right) +\mathcal{O}(\e^2,\r_{min}^0)\;.
}
The numerical factor $\frac{1}{4}$ can be grouped with the rest of the finite $\r_{min}^0$ terms and will be ignored for now.  Finally using the result of Brown and Henneaux \cite{Brown:1986nw}
\eqn{BRHENN}{
c= \frac{3 \ell}{2G_{N}} \;,
}
\eno{GOTSBULK} evaluates to 
\eqn{GOTSBULKC}{
S_{bulk} =  -\frac{c}{12\pi} \int_{\p \mathcal{N}} dz d\bar{z} \e G^{1,1}(z,\bar{z})\log\frac{\r_{max}}{\r_{min}}+\mathcal{O}(\e^2,\r_{min}^0)\;.
}
Of course, alternatively one can use the result of \cite{Henningson:1998gx}, whose derivation of \eno{BRHENN} this paper closely follows.  Finally we need to use $G^{1,1}(z,\bar{z})=-2\pi \d^{2}(z,\bar{z})$ and that the cutoff $\r_{min}$ is related to the usual UV cutoff with 
\eqn{RMIN}{
\r_{min} = \alpha^2 
}
to derive 
\eqn{GOTEEFIN}{
S^{E.E.}_{hol.}= \frac{c}{3}\log\frac{\sqrt{\r_{max}}}{\a}+finite\;.
}
For the case of pure $AdS_{3}$ it is reasonable to take $\sqrt{\r_{max}}\sim L$. For the non-rotating BTZ black hole, we just integrate up to the location of the horizon $\sqrt{r_{max}}=\frac{1}{r_{+}}\sim \b$, see appendix \eno{BTZ}.  When the temperature is larger than the periodicity of the angular coordinate $\theta$, $l$,  which is the case when the BTZ solution should be used, then these both reproduce \eno{EESINGLE}-\eno{EEFINITET}.  It should be noted that the procedure followed in this paper does not separate between the different finite parts, as is the usual case in other computations \cite{Calabrese:2004eu,Ryu:2006bv} 
\eqn{SEEFINITE}{
S^{E.E.} = \frac{c}{3}\left(\log \frac{1}{\a} +\log L \right) +finite, \quad S^{E.E.} = \frac{c}{3}\left(\log \frac{1}{\a} + \log \frac{\b}{\pi}\sinh(\frac{\pi l}{\b})\right) +finite \;,
} 
for the cases of pure AdS and the BTZ black hole.  

We should note here that currently there is no rigorous way to find $\r_{max.}$.  One would hope that it will come directly from some requirement, such as regularity for the solution for the bulk metric.  Finding the proper prescription for determining $\r_{max}$ does not seem possible in the current set up, since the ``uniformisation`` process has sent the boundary points of $A$ from $u=0,v=L$ to $0$ and $\infty$ respectively. Once $\r_{max}$ is properly determined, the finite part of entanglement entropy will also be known and calculable. As far as finite parts of the entanglement entropy are concerned, holographic renormalization should also be taken into account, as counterterms will contribute to the finite part of the bulk action.

\section{Conclusions and discussion}\label{DISC}

In this note we have examined a straightforward approach to computing the entanglement entropy holographically.  For two dimensions the known results are reproduced.  However the most interesting cases are the higher dimensional ones.  In view of the results of \cite{Schwimmer:2008yh}, the prescription of \cite{Ryu:2006bv,Ryu:2006ef} does not have the same analytic structure as the boundary theory predicts.  It could be that that prescription describes some other Wilson loop type of observable.  Another possibility is that the replica trick fails in higher dimensions.  One could calculate the entanglement entropy, with and without the replica trick and compare them.  Unfortunately even for the simple case of a free boson and a sphere, it appears that only a numerical computation is possible.

A way in which the replica trick could fail in higher dimensions is that there is no analytic continuation from integer to real values of $n$.  If we have a massless boson in four dimensions, we can transform the evaluation of the entanglement entropy by dimensionally reducing the theory to two dimensions.  Then, the entanglement entropy of an infinite tower of massive bosons has to be computed.  We already know that the calculation  for certain two dimensional massive theories involves a highly non trivial analytic continuation.  Indeed,  some massive two dimensional quantum integrable systems were examined in \cite{ Cardy:2007mb}.  The analytic structure of the two point funtion of the energy momentum tensor was examined and it was found that there is no natural analytic continuation from $n=1,2,3...$ to $[1,\infty)$.  However there is a unique analytic continuation from $n=2,3,...$ to $[1,\infty)$, if certain assumptions for the behavior at infinity are made.  Interestingly, in those models kinematic singularities also contribute to the analytic structure of the two point function.  One would expect that similar treatment for the analyticity properties has to be followed for the higher dimensional theories.   

Another way to try to derive the Ryu-Takayanagi proposal would be the following.  When the switch to the singular metric is done one adds to the boundary metric a localized small perturbation $\e g_{sing.,ij}$.  This would amount to adding a localized stress-energy tensor operator $T_{ij}$ to the boundary.  Then the partition function will have to be calculated with an insertion of this operator.  This is reminiscent of the method used in \cite{Calabrese:2004eu,Calabrese:2005zw} for $d=2$, but it is unclear to me how to generalize this to an arbitrary higher dimension. 

The next step is therefore to examine how to implement this direct approach to calculating the entanglement entropy in higher dimensions.  Generically one does not expect the Taylor series of $g_{ij}$ in terms of $\r$ to terminate at a finite number of steps, see for example the approach of \cite{deHaro:2000xn}.  Adding a small singular term to the boundary metric complicates things.  As the case of $d=2$, one would expect the higher order terms to be derivatives of the lower order ones and that will produce an increasingly more singular behavior in the metric.  However certain simple situations like a sphere or a cylinder on an non thermal background should have a simple answer.  In a certain sense, in two dimensions one is bound to find the correct answer, since the result is governed by the conformal anomaly.  In higher dimensions, where the structure of the anomalies is much richer, it is uncertain whether a given holographic prescription calculates the quantity with the correct analytic structure.  The advantage of the prescription followed in this note is that the analytic structure of the CFT quantity and the holographic answer are the same by construction.  

Another interesting aspect of entanglement entropy has to do with black holes.  It has been argued that in some cases the entropy of the black hole is the entanglement entropy between the states leaving inside and outside the horizon, see \cite{Hawking:2000da} for a cosmological horizon example and \cite{Emparan:2006ni} for a black hole in a Randall-Sundrum scenario.  Certain two dimensional models exhibit many of the interesting features of black hole formation, Hawking radiation and so on, for a review see \cite{Strominger:1994tn}.  It would be intriguing to find a holographic dual for these two dimensional models.  We leave some of these very interesting questions for future work.

\section{Acknowledgements}\label{ACKN}
I would like to thank C.\;Bachas, A.\;Schwimmer and K.\;Sfetsos for discussions, and E.Kiritsis for comments on an early draft.  The author is supported by ANR grant  ANR-05-BLAN-0079-02. This work was also 
partially supported by  RTN contracts MRTN-CT-2004-005104 and
MRTN-CT-2004-503369, CNRS PICS \#~2530,  3059 and 3747,
 and by a European Union Excellence Grant, MEXT-CT-2003-509661.

\clearpage
\appendix

\section{The (non-rotating) BTZ black hole in Fefferman-Graham coordinates}\label{BTZ}

In this appendix we derive the BTZ black hole using the methods of \cite{Skenderis:1999nb}.  One only has to assume that the asymptotic boundary metric is of the form 
\eqn{BTZbound}{
ds^2 = (d\t^2+d\theta^2)\;.
}
Then the equation that needs to be satisfied by the scalar $\f$ is 
\eqn{BTZboundCurv}{
\Box \f= R_{0} =0
}
with a general solution 
\eqn{GOTBTZscalar}{
\f = A \th +B \t + \G\;.
}
We take the periodicity in the $\th$ coordinate to be large, so as not to worry about boundary conditions.  For the same reason lets also choose $B=0$.  $\G$ has no effect on the solution and can be conveniently dropped.  It is straightforward to write the second component of the metric as 
\eqn{GTWOBTZ}{
g_{2} = \left(\begin{array}{cc} \ell^2 A^2/8 & 0\\ 0 &\ell^2 A^2/8 \end{array} \right)\;.
} 
Then. the prescription gives the metric
\eqn{GOTBTZMETRIC}{
ds^2 = \frac{\ell^2d \r ^2}{4 \r^2} + \frac{\left(\r(\ell/2A)^2+1\right)^2d\th^2+(\r(\ell/2A)^2-1)2d\t^2}{4\r}\;.
}
With the identification
\eqn{GOTA}{
A = 2r_{+}/\ell}
and the change of coordinates to 
\eqn{BTZNEWr}{
\r = \left(r+\sqrt{r^2-r_{+}^{2}}\right)^{-2}
}
the metric is brought to the familiar form 
\eqn{BTZMETRIC}{
ds^{2} = \frac{\ell^2dr^2}{r^{2}-r_{+}^2}+\left(r^2-r_{+}^2\right)d\t^2+r^2d\th^2\;.
}
The $\r$ coordinate takes values between 
\eqn{RANGER}{
\r \in (0,\r_{+})\;,
}
where $\r_{+}$ is the location of the horizon 
\eqn{GOTRRMAX}{
\r_{+} =\frac{1}{r_{+}^{2}}\;.
}
Keeping also a non zero B, one can reproduce the rotating black hole solution.

\clearpage
\bibliographystyle{xbib}
\bibliography{eeanomal}

\providecommand{\href}[2]{#2}\begingroup\raggedright\begin{thebibliography}{10}

\bibitem{Ryu:2006bv}
S.~Ryu and T.~Takayanagi, ``{Holographic derivation of entanglement entropy
  from AdS/CFT},'' {\em Phys. Rev. Lett.} {\bf 96} (2006) 181602,
\href{http://arXiv.org/abs/hep-th/0603001}{{\tt hep-th/0603001}}.

\bibitem{Ryu:2006ef}
S.~Ryu and T.~Takayanagi, ``{Aspects of holographic entanglement entropy},''
  {\em JHEP} {\bf 08} (2006) 045,
\href{http://arXiv.org/abs/hep-th/0605073}{{\tt hep-th/0605073}}.

\bibitem{Schwimmer:2008yh}
A.~Schwimmer and S.~Theisen, ``{Entanglement Entropy, Trace Anomalies and
  Holography},''
\href{http://arXiv.org/abs/0802.1017}{{\tt 0802.1017}}.

\bibitem{Solodukhin:2008dh}
S.~N. Solodukhin, ``{Entanglement entropy, conformal invariance and extrinsic
  geometry},''
\href{http://arXiv.org/abs/0802.3117}{{\tt 0802.3117}}.

\bibitem{Callan:1994py}
J.~Callan, Curtis~G. and F.~Wilczek, ``{On geometric entropy},'' {\em Phys.
  Lett.} {\bf B333} (1994) 55--61,
\href{http://arXiv.org/abs/hep-th/9401072}{{\tt hep-th/9401072}}.

\bibitem{Calabrese:2004eu}
P.~Calabrese and J.~L. Cardy, ``{Entanglement entropy and quantum field
  theory},'' {\em J. Stat. Mech.} {\bf 0406} (2004) P002,
\href{http://arXiv.org/abs/hep-th/0405152}{{\tt hep-th/0405152}}.

\bibitem{Calabrese:2005zw}
P.~Calabrese and J.~L. Cardy, ``{Entanglement entropy and quantum field theory:
  A non- technical introduction},'' {\em Int. J. Quant. Inf.} {\bf 4} (2006)
  429,
\href{http://arXiv.org/abs/quant-ph/0505193}{{\tt quant-ph/0505193}}.

\bibitem{Holzhey:1994we}
C.~Holzhey, F.~Larsen, and F.~Wilczek, ``{Geometric and renormalized entropy in
  conformal field theory},'' {\em Nucl. Phys.} {\bf B424} (1994) 443--467,
\href{http://arXiv.org/abs/hep-th/9403108}{{\tt hep-th/9403108}}.

\bibitem{Zamolodchikov:1987ti}
A.~B. Zamolodchikov, ``{Renormalization Group and Perturbation Theory Near
  Fixed Points in Two-Dimensional Field Theory},'' {\em Sov. J. Nucl. Phys.}
  {\bf 46} (1987)
1090.

\bibitem{Casini:2004bw}
H.~Casini and M.~Huerta, ``{A finite entanglement entropy and the c-theorem},''
  {\em Phys. Lett.} {\bf B600} (2004) 142--150,
\href{http://arXiv.org/abs/hep-th/0405111}{{\tt hep-th/0405111}}.

\bibitem{Casini:2003ix}
H.~Casini, ``{Geometric entropy, area, and strong subadditivity},'' {\em Class.
  Quant. Grav.} {\bf 21} (2004) 2351--2378,
\href{http://arXiv.org/abs/hep-th/0312238}{{\tt hep-th/0312238}}.

\bibitem{Maldacena:1997re}
J.~M. Maldacena, ``{The large N limit of superconformal field theories and
  supergravity},'' {\em Adv. Theor. Math. Phys.} {\bf 2} (1998) 231--252,
\href{http://arXiv.org/abs/hep-th/9711200}{{\tt hep-th/9711200}}.

\bibitem{Gubser:1998bc}
S.~S. Gubser, I.~R. Klebanov, and A.~M. Polyakov, ``{Gauge theory correlators
  from non-critical string theory},'' {\em Phys. Lett.} {\bf B428} (1998)
  105--114,
\href{http://arXiv.org/abs/hep-th/9802109}{{\tt hep-th/9802109}}.

\bibitem{Witten:1998qj}
E.~Witten, ``{Anti-de Sitter space and holography},'' {\em Adv. Theor. Math.
  Phys.} {\bf 2} (1998) 253--291,
\href{http://arXiv.org/abs/hep-th/9802150}{{\tt hep-th/9802150}}.

\bibitem{Aharony:1999ti}
O.~Aharony, S.~S. Gubser, J.~M. Maldacena, H.~Ooguri, and Y.~Oz, ``{Large N
  field theories, string theory and gravity},'' {\em Phys. Rept.} {\bf 323}
  (2000) 183--386,
\href{http://arXiv.org/abs/hep-th/9905111}{{\tt hep-th/9905111}}.

\bibitem{Fursaev:2006ih}
D.~V. Fursaev, ``{Proof of the holographic formula for entanglement entropy},''
  {\em JHEP} {\bf 09} (2006) 018,
\href{http://arXiv.org/abs/hep-th/0606184}{{\tt hep-th/0606184}}.

\bibitem{Skenderis:1999nb}
K.~Skenderis and S.~N. Solodukhin, ``{Quantum effective action from the AdS/CFT
  correspondence},'' {\em Phys. Lett.} {\bf B472} (2000) 316--322,
\href{http://arXiv.org/abs/hep-th/9910023}{{\tt hep-th/9910023}}.

\bibitem{deHaro:2000xn}
S.~de~Haro, S.~N. Solodukhin, and K.~Skenderis, ``{Holographic reconstruction
  of spacetime and renormalization in the AdS/CFT correspondence},'' {\em
  Commun. Math. Phys.} {\bf 217} (2001) 595--622,
\href{http://arXiv.org/abs/hep-th/0002230}{{\tt hep-th/0002230}}.

\bibitem{GrahamFeff}
C.Fefferman and C.R.Graham, ``{Conformal Invariants},'' {\em Ellie Cartan et
  les Mathematiques d'ajourd'hui} (Asterisk,1985) 95.

\bibitem{Nishioka:2006gr}
T.~Nishioka and T.~Takayanagi, ``{AdS bubbles, entropy and closed string
  tachyons},'' {\em JHEP} {\bf 01} (2007) 090,
\href{http://arXiv.org/abs/hep-th/0611035}{{\tt hep-th/0611035}}.

\bibitem{Klebanov:2007ws}
I.~R. Klebanov, D.~Kutasov, and A.~Murugan, ``{Entanglement as a Probe of
  Confinement},'' {\em Nucl. Phys.} {\bf B796} (2008) 274--293,
\href{http://arXiv.org/abs/0709.2140}{{\tt 0709.2140}}.

\bibitem{Azeyanagi:2007bj}
T.~Azeyanagi, T.~Nishioka, and T.~Takayanagi, ``{Near Extremal Black Hole
  Entropy as Entanglement Entropy via AdS2/CFT1},'' {\em Phys. Rev.} {\bf D77}
  (2008) 064005,
\href{http://arXiv.org/abs/0710.2956}{{\tt 0710.2956}}.

\bibitem{Hubeny:2007re}
V.~E. Hubeny and M.~Rangamani, ``{Holographic entanglement entropy for
  disconnected regions},'' {\em JHEP} {\bf 03} (2008) 006,
\href{http://arXiv.org/abs/0711.4118}{{\tt 0711.4118}}.

\bibitem{Hubeny:2007xt}
V.~E. Hubeny, M.~Rangamani, and T.~Takayanagi, ``{A covariant holographic
  entanglement entropy proposal},'' {\em JHEP} {\bf 07} (2007) 062,
\href{http://arXiv.org/abs/0705.0016}{{\tt 0705.0016}}.

\bibitem{Pakman:2008ui}
A.~Pakman and A.~Parnachev, ``{Topological Entanglement Entropy and
  Holography},''
\href{http://arXiv.org/abs/0805.1891}{{\tt 0805.1891}}.

\bibitem{Barbon:2008ut}
J.~L.~F. Barbon and C.~A. Fuertes, ``{Holographic entanglement entropy probes
  (non)locality},'' {\em JHEP} {\bf 04} (2008) 096,
\href{http://arXiv.org/abs/0803.1928}{{\tt 0803.1928}}.

\bibitem{Azeyanagi:2007qj}
T.~Azeyanagi, A.~Karch, T.~Takayanagi, and E.~G. Thompson, ``{Holographic
  Calculation of Boundary Entropy},'' {\em JHEP} {\bf 03} (2008) 054--054,
\href{http://arXiv.org/abs/0712.1850}{{\tt 0712.1850}}.

\bibitem{Headrick:2007km}
M.~Headrick and T.~Takayanagi, ``{A holographic proof of the strong
  subadditivity of entanglement entropy},'' {\em Phys. Rev.} {\bf D76} (2007)
  106013,
\href{http://arXiv.org/abs/0704.3719}{{\tt 0704.3719}}.

\bibitem{Cadoni:2007vf}
M.~Cadoni, ``{Entanglement entropy of two-dimensional Anti-de Sitter black
  holes},'' {\em Phys. Lett.} {\bf B653} (2007) 434--438,
\href{http://arXiv.org/abs/0704.0140}{{\tt 0704.0140}}.

\bibitem{Hirata:2006jx}
T.~Hirata and T.~Takayanagi, ``{AdS/CFT and strong subadditivity of
  entanglement entropy},'' {\em JHEP} {\bf 02} (2007) 042,
\href{http://arXiv.org/abs/hep-th/0608213}{{\tt hep-th/0608213}}.

\bibitem{Calabrese:2005in}
P.~Calabrese and J.~L. Cardy, ``{Evolution of Entanglement Entropy in
  One-Dimensional Systems},'' {\em J. Stat. Mech.} {\bf 0504} (2005) P010,
\href{http://arXiv.org/abs/cond-mat/0503393}{{\tt cond-mat/0503393}}.

\bibitem{Tsallis:1987eu}
C.~Tsallis, ``{Possible Generalization of Boltzmann-Gibbs Statistics},'' {\em
  J. Stat. Phys.} {\bf 52} (1988)
479--487.

\bibitem{Brown:1986nw}
J.~D. Brown and M.~Henneaux, ``{Central Charges in the Canonical Realization of
  Asymptotic Symmetries: An Example from Three-Dimensional Gravity},'' {\em
  Commun. Math. Phys.} {\bf 104} (1986)
207--226.

\bibitem{Henningson:1998gx}
M.~Henningson and K.~Skenderis, ``{The holographic Weyl anomaly},'' {\em JHEP}
  {\bf 07} (1998) 023,
\href{http://arXiv.org/abs/hep-th/9806087}{{\tt hep-th/9806087}}.

\bibitem{Cardy:2007mb}
J.~L. Cardy, O.~A. Castro-Alvaredo, and B.~Doyon, ``{Form factors of
  branch-point twist fields in quantum integrable models and entanglement
  entropy},''
\href{http://arXiv.org/abs/0706.3384}{{\tt 0706.3384}}.

\bibitem{Hawking:2000da}
S.~Hawking, J.~M. Maldacena, and A.~Strominger, ``{DeSitter entropy, quantum
  entanglement and AdS/CFT},'' {\em JHEP} {\bf 05} (2001) 001,
\href{http://arXiv.org/abs/hep-th/0002145}{{\tt hep-th/0002145}}.

\bibitem{Emparan:2006ni}
R.~Emparan, ``{Black hole entropy as entanglement entropy: A holographic
  derivation},'' {\em JHEP} {\bf 06} (2006) 012,
\href{http://arXiv.org/abs/hep-th/0603081}{{\tt hep-th/0603081}}.

\bibitem{Strominger:1994tn}
A.~Strominger, ``{Les Houches lectures on black holes},''
\href{http://arXiv.org/abs/hep-th/9501071}{{\tt hep-th/9501071}}.

\end{thebibliography}\endgroup

\end{document}